%% file: monr2water.tex
\documentclass[structuredabstract]{aa}  
\usepackage{graphicx}
\usepackage{natbib}
\usepackage{xspace}
\xspaceaddexceptions{]\}}

\usepackage[colorlinks, linkcolor=blue, citecolor=blue, breaklinks=true]{hyperref}
\usepackage[update,prepend]{epstopdf}
\usepackage{booktabs}
\usepackage{amsmath}

\newcommand{\rarrow}{\ensuremath{\rightarrow}}

\newcommand{\ex}[2]{\ensuremath{#1 \times 10^{#2}}\xspace}

\newcommand{\kms}{km\,s$^{-1}$\xspace}
\newcommand{\cmc}{cm$^{-3}$\xspace}
\newcommand{\cmq}{cm$^{-2}$\xspace}

\newcommand{\um}{$\mu$m\xspace}

\newcommand{\CII}{\ion{C}{ii}\xspace}

\newcommand{\HII}{\ion{H}{ii}\xspace}

\newcommand{\hd}{H$_2$\xspace}
\newcommand{\wat}{H$_2$O\xspace}
\newcommand{\amm}{NH$_3$\xspace}
\newcommand{\dco}{$^{12}$CO\xspace}
\newcommand{\cdo}{C$^{18}$O\xspace}
\newcommand{\tco}{$^{13}$CO\xspace}

\newcommand{\GILDAS}{\texttt{GILDAS}\xspace}
\newcommand{\CLASS}{\texttt{CLASS}\xspace}

\newcommand{\herschel}{{\it Herschel}\xspace}

\usepackage{txfonts}

\begin{document}
\title{\herschel\ / HIFI observations of CO, \wat and \amm in Mon~R2\thanks{\herschel is an ESA space observatory with science instruments provided by European-led Principal Investigator consortia and with important participation from NASA.}\fnmsep\thanks{Based on observations carried out with the IRAM 30m Telescope. IRAM is supported by INSU/CNRS (France), MPG (Germany) and IGN (Spain).}}

\input{authors.tex}
\authorrunning {P. Pilleri, et al.} 
\titlerunning{\herschel\ / HIFI observations of CO, \wat and \amm in Mon~R2}


\abstract
{Mon R2, at a distance of 830 pc, is the only ultracompact \HII\ region (UC\HII)  where the associated photon-dominated region (PDR) can be resolved with {\it Herschel}. Owing to its brightness and proximity, it is one of the best-suited sources for investigating the chemistry and physics of highly UV-irradiated PDRs.}
{
Our goal is to  estimate the abundance of  \wat and \amm  in this region and investigate their origin.
}
{
We present new observations  ([\CII], \dco, \tco, \cdo, o-\wat, p-\wat, o-H$_2^{18}$O
and o-\amm)  obtained with the HIFI instrument onboard \herschel and  the IRAM-30\,m telescope. 
We investigated the physical conditions in which these lines arise by analyzing their velocity structure  and spatial variations. Using 
a large velocity gradient approach, we modeled the line intensities and derived an average abundance of \wat and \amm across the region. Finally, 
we modeled the line profiles with a non-local radiative transfer model and compared these results with the  abundance predicted by the Meudon PDR  code.
 }
{
The  variations of the line profiles and intensities indicate complex geometrical and kinematical patterns. 
In several tracers ([\CII], CO 9\rarrow8 and \wat) the line profiles vary significantly with position and have broader line widths toward the \HII region.  The \wat lines present 
 strong self-absorption at the ambient velocity and emission in high-velocity wings toward the \HII region. The emission in the o-H$_2^{18}$O ground state line reaches its maximum value around 
the \HII region, has smaller linewidths and peaks at the velocity of the ambient cloud. Its spatial distribution shows that the o-H$_2^{18}$O  emission arises in 
the  PDR surrounding the \HII region. By modeling the o-H$_2^{18}$O  emission and assuming the standard $[^{16}$O$]/[^{18}$O$]=500$, we derive a 
mean abundance of o-\wat  of $\sim 10^{-8}$ relative to H$_2$. The ortho-\wat  abundance, however, is larger ($\sim 1\times10^{-7}$) in the high-velocity wings detected
toward the \HII region.  Possible explanations for this larger abundance include an expanding  hot PDR and/or an outflow. Ammonia seems to be present only in the envelope of the core with an average abundance of $\sim \ex{2}{-9}$ relative to \hd.
 }
{The Meudon PDR code, which includes only gas-phase chemical networks, can  account for the measured water abundance
in the high velocity gas as long as we assume that it originates from a $\lesssim$1~mag hot expanding layer of the PDR, i.e. that the outflow has only a minor contribution to this emission. To explain the water and 
ammonia abundances in the rest of the cloud,  the molecular freeze out and grain surface chemistry would need to be included.}
 

\maketitle

\section{Introduction}

\begin{table}
\caption{Summary of HIFI and 30m observations}
\label{table_obssum}
\begin{center}
\begin{tabular}{lcccc}
\toprule
Line		& $\nu$	    & HPBM  & $\eta_l$ \tablefootmark{a}& Telescope  \\
				& [GHz]	    & [\arcsec] &	&	           \\
\midrule
$^{12}$CO  (9$\rarrow$8)  		& 1036.912  & 20.4  & 0.77	& \herschel    \\
\\
$^{13}$CO 	(2$\rarrow$1)		& 220.398   & 10.0  & 0.67	& IRAM      \\
$^{13}$CO 	(5$\rarrow$4)  		& 550.926   & 38.5  & 0.79   	& \herschel      \\
$^{13}$CO 	(10$\rarrow$9) 	& 1101.349   & 19.3 & 0.77   	& \herschel       \\
\\	
C$^{18}$O 	(2$\rarrow$1)  		& 219.560   & 10.0  & 0.67	& IRAM      \\
C$^{18}$O 	(5$\rarrow$4)  		& 548.830   & 38.6  & 0.79   	& \herschel       \\
\\
o$-$H$_2$O      (1$_{10}\rarrow1_{01}$) & 556.936   & 38.1  & 0.79 	& \herschel      \\
p$-$H$_2$O      (1$_{11}\rarrow0_{00}$) & 1113.343  & 19.3  & 0.77 	& \herschel       \\
o$-$H$_2^{18}$O  (1$_{10}\rarrow1_{01}$) & 547.676   & 38.0  & 0.79 	& \herschel       \\
\\
o-NH$_3$ (1$_{0}\rarrow0_{0}$) 	& 572.498   & 37.0  & 0.79 	& \herschel      \\
\\
H	(42$\alpha$)  		& 85.688   & 29  & 0.63	& IRAM       \\
\\
$[$\CII$]$		($^2P_{3/2}\rarrow^2P_{1/2}$)		& 1900.537  & 11.2  & 0.73       & \herschel    \\
\bottomrule
\end{tabular}
\end{center}
\tablefoottext{a}{$\eta_l = B_{\rm{eff}}/F_{\rm{eff}}$}. \\
\end{table}

Although the processes that lead to the formation of massive stars are still not fully understood, it is generally agreed that massive stellar objects are formed by the collapse of a dense molecular cloud into one or multiple self-gravitating pre-stellar objects. Once the star is born, the innermost layers of the molecular cloud are heated and ionized by the strong UV radiation field, producing what 
is called an ultra compact \HII region (UC\HII). 
These regions are characterized by extreme
UV irradiation and very small physical scales ($\lesssim0.1$~pc), and are
embedded in dense molecular clouds with gas densities often higher than 10$^6$~cm$^{-3}$ \citep{hoare07}. 
Whereas the H-ionizing radiation is absorbed in  the \HII region, UV radiation carrying energies less than 13.6~eV penetrates 
deeper into the molecular cloud, producing a so-called photo-dissociation region (PDR).  In these regions, the chemistry and the physics are driven by the extreme impinging radiation field \citep[more than $10^5$ times the Habing field $G_{\rm 0}$, see ][]{habing68}.
The study of UC \HII regions is crucial to understanding the different processes in massive star formation.

\begin{figure*}
\centering
\includegraphics[width=0.47\linewidth]{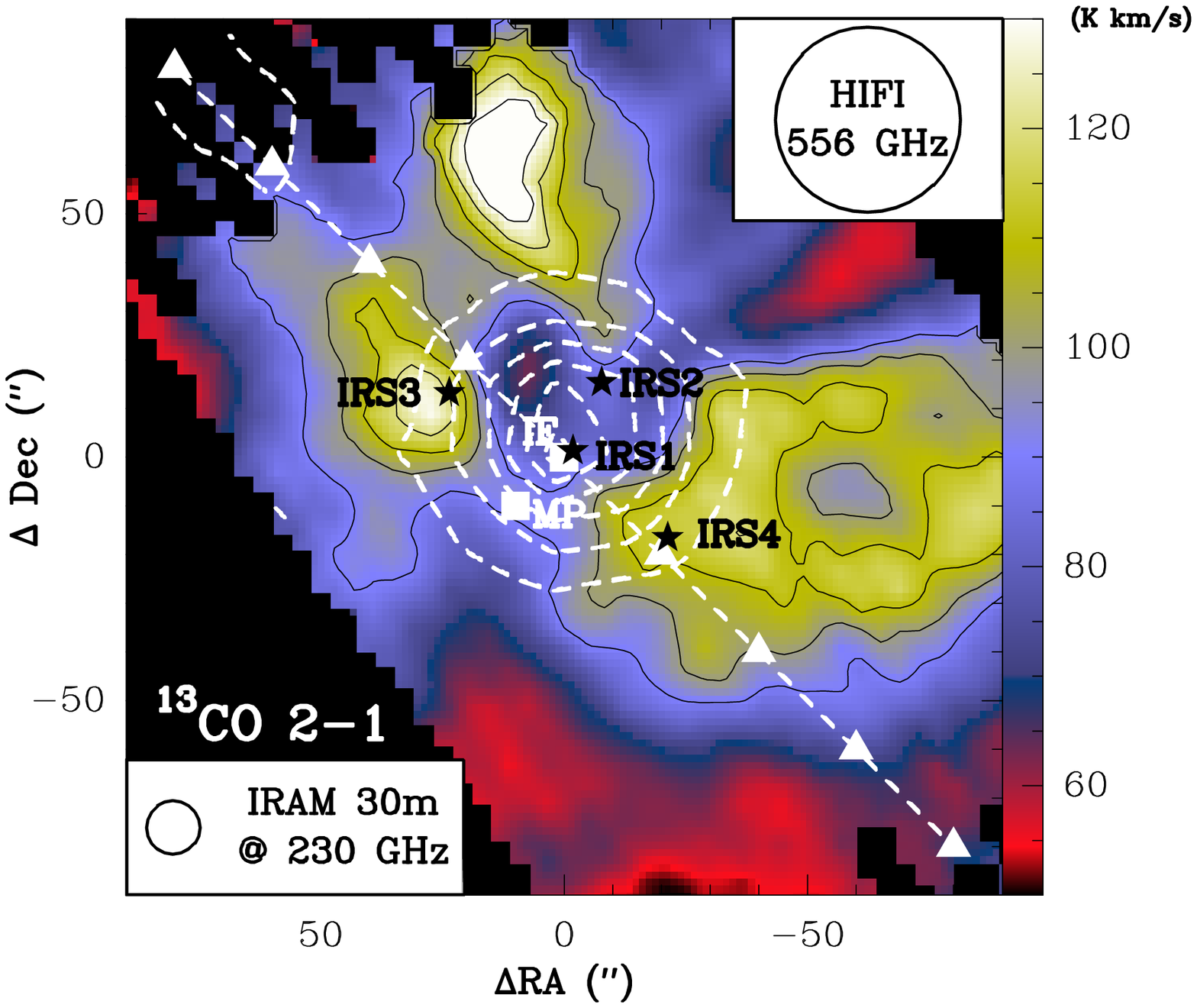}
\includegraphics[width=0.47\linewidth]{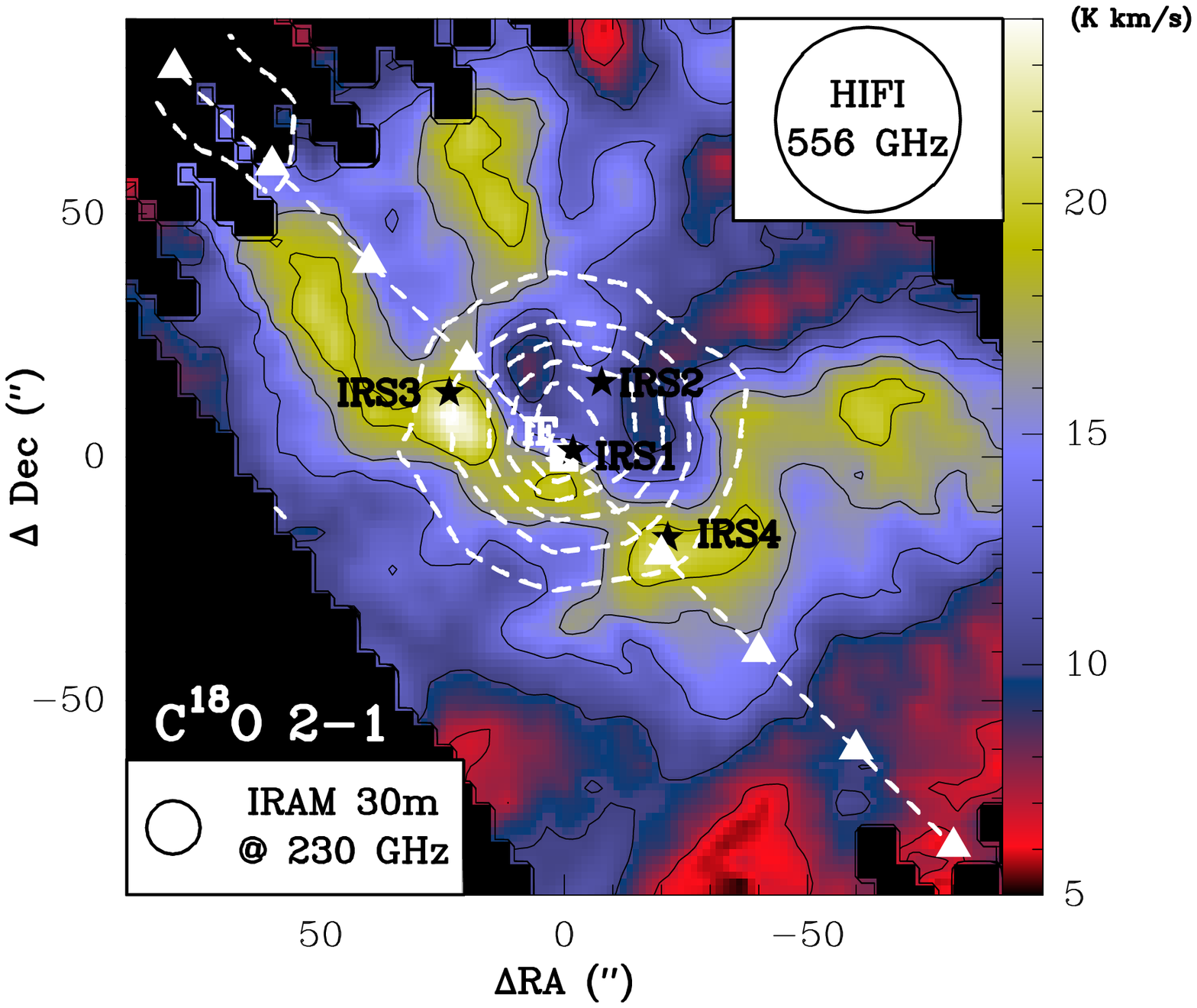}
\caption{Maps of the integrated intensity  (color scale)  between 5 and 15~\kms of 
the \tco (2\rarrow1, left) and \cdo (2\rarrow1, right) lines observed 
at the IRAM-30m telescope. Dashed white contours represent the integrated 
intensity of the H42$\alpha$ recombination line at 85.688 GHz (1 to 11~K~\kms 
in steps of 2~K~\kms), tracing the \HII region. Squares represent 
the positions of the ionization front (IF) and the molecular peak (MP), whereas the triangles represent 
the points of the observed strip (dashed line) for \dco (9\rarrow8), 
\wat and \amm that are studied in this work. The infrared sources are indicated with black stars.}
\label{fig_mapmon2}
\end{figure*}

Mon R2 is the UC \HII region created by a B0 star \citep{downes75} within the nearby (830\,pc) star-forming region Monoceros. It  is the closest UC \HII region and, with an angular diameter of $\sim$22\arcsec (0.1 pc), the only one that can be resolved by single-dish millimeter and 
far-IR telescopes.
 The formation of the star associated to the infrared source IRS1 created
a huge bipolar outflow \citep[$\sim$15$\arcmin$ = 3.6 pc long, ][]{massi85,henning92,tafalla94}, which is  now inactive. A compact active bipolar outflow, more likely related to FIRS~3,  was observed  in the low-J CO \citep{giannakopoulou97} and methanol \citep{xu06} lines.
The UC \HII region  has
a cometary shape and reaches its maximum continuum brightness toward the  infrared source
Mon R2 IRS 1.
The host molecular cloud has been characterized by many previous millimeter spectroscopic and continuum studies
 \citep{henning92, giannakopoulou97, tafalla97, choi00, rizzo03, rizzo05}. The molecular emission shows an arclike structure
surrounding the \HII region, with the bulk of the emission to the southwest (see Fig. \ref{fig_mapmon2}). 
  The UC\HII appears to be opened to the north, as shown by the very extended emission of small dust particles at 8\,$\mu$m, in a region characterized by a PDR-like chemistry \citep{ginard11}.
Depending on the tracer, gas densities  $n_{\rm{H}_2}$ between a few 10$^5$\,\cmc and
$\sim 5 \times 10^6$\,\cmc have been determined in the molecular  gas,  which testifies the presence of  large gradients in the gas densities of this region \citep{rizzo03, berne09a, ginard11}.
Recent high spatial resolution observations in the mid-infrared of the  \hd  rotational lines and of UV-excited polycyclic aromatic hydrocarbons (PAHs)
have shown the existence of a thin  layer 
 of hot ($T_{\rm k}$=100-600~K) and relatively dense molecular gas  ($n_{{\rm H}_2} \sim 10^5$ cm$^{-3}$, $N$({\rm H}$_2$)$\sim1\times$10$^{21}$~cm$^{-2}$)  in between 
the ionized region and the dense molecular cloud \citep{berne09a}. 
The detection of the reactive ions CO$^+$ and HOC$^+$ toward FIRS 1 using the IRAM-30m telescope constitute additional  proof for the existence of
a  high-density PDR  \citep{rizzo03}.

Mon~R2 has been targeted as part of the \herschel guaranteed-time key program ``Warm and Dense Interstellar Medium" (WADI, PI: V. Ossenkopf) as prototype of UC \HII regions.
The first \herschel observations of this source were presented by \citet{fuente10}, who reported the detection of several far-IR/sub-mm lines,  including o-H$_2$$^{18}$O 1$_{1,1}$$\rightarrow$0$_{0,0}$, 
toward the so-called molecular peak (hereafter MP, see Fig. \ref{fig_mapmon2}). On the basis of these observations they estimated an average water abundance across the PDR of $\sim2\times 10^{-8}$ relative to \hd.  This value is slightly higher than that obtained by \citet{snell00}, which was based on lower angular resolution  observations of the o-\wat ground state line  with SWAS ($\sim \ex{2}{-9}$ relative to \hd).

In this paper we  present new \herschel observations and  complementary mm data obtained at the IRAM-30m telescope
to improve our knowledge of the molecular gas surrounding the \HII region, and investigate the origin of the CO, \wat and \amm emission in more details.  

\section{Observations}
\label{sec_observations}

Table \ref{table_obssum} 
summarizes the observed transitions with the
corresponding frequencies, beam sizes, and beam efficiencies\footnote{http://herschel.esac.esa.int/Docs/TechnicalNotes/HIFI\_Beam\_Effi\-ciencies\_17Nov2010.pdf}$^,$\footnote{http://www.iram.es/IRAMES/mainWiki/Iram30mEfficiencies}. 
In this paper, the intensity scale is main beam temperature ($T_{\rm{mb}}$), 
and offsets were calculated relative to the ionization 
front (hereafter IF: RA$_{J2000}$=06$^h$07$^m$46.2$^s$, DEC$_{J2000}$=-06\degr23$'$08.3$''$).

\subsection{\herschel observations}

The \herschel \citep{pilbratt10} observations presented here were obtained with 
the heterodyne instrument for the far infrared HIFI \citep[][]{degraauw10} 
as part of the WADI guaranteed-time key program \citep{ossenkopf11}. 
A single local oscillator (LO) configuration allowed us to simultaneously observe the o-\wat 
($1_{10}\rarrow1_{01}$) and o-\amm (1$_{0}\rarrow0_{0}$) lines. 
A second setup was dedicated to the observation of the p-\wat 
($1_{11}\rarrow1_{00}$) and \tco (10\rarrow9) lines.
 A third LO setting allowed us to observe the o-H$_2^{18}$O ($1_{10}\rarrow1_{01}$) line as well as the \tco (5\rarrow4) and \cdo (5\rarrow4) lines. Finally, dedicated setups were used for the \dco (9\rarrow8) line 
and the [\CII] line.
 All  lines were observed in one strip across the region, oriented 45$^\circ$ (east of north). The [\CII], o-\wat, p-\wat and \dco strips extend 2\arcmin\, in each direction relative to the IF (see Fig.~\ref{fig_mapmon2}). To achieve a better signal-to-noise ratio (S/N) in  the o-H$_2^{18}$O ($1_{10}\rarrow1_{01}$) line, this strip extended for only $\sim$ 1\arcmin\ on each side of the IF.
The strips were obtained using the on-the-fly (OTF) observing mode using Nyquist sampling and  with the reference position  at the offset  (+10\arcmin; 0\arcmin), which is free of emission.

 The basic data reduction was performed using the
standard pipeline provided with the 
version 7.0 of HIPE\footnote{HIPE is a joint development by the Herschel 
Science Ground Segment Consortium, consisting of ESA, the NASA Herschel 
Science Center, and the HIFI, PACS and SPIRE consortia} \citep{ott10} 
and then exported to \GILDAS/\CLASS \citep{pety05} for a more detailed analysis. 
  For the \tco (10\rarrow9) and p-\wat ($1_{11}\rarrow1_{00}$) lines we subtracted a secondary OFF spectrum obtained at a position located at the NE end of the strip that is free of emission. Typical noise {\it rms} values are $\sim $ 20~mK 
(o-\wat and o-\amm), $\sim 0.2$~K (\dco), $\sim $ 20~mK (\tco 10\rarrow9 and p-\wat), 7~mK (o-H$_2^{18}$O, \tco 5\rarrow4 and \cdo 5\rarrow4) 
and 1~K ([\CII]), all calculated at the resolution of 0.7~\kms. 

Because HIFI is a double-sideband receiver, but all our spectra are calibrated to the single-sideband scale, the continuum levels need to be corrected for the contribution from the image sideband. For all our spectra, we can assume a sideband gain ratio of unity \citep{roelfsema12}, so that the continuum level needs to be divided by a factor two.
 After this correction, the continuum levels toward the IF are 0.25 and 0.5\,K at 550\,GHz and 1110\,GHz, respectively. 
The [\CII] observations showed strong
standing waves, which were removed using the {\it FitHifiFringe} task within HIPE.    These standing waves hinder the determination of the continuum level at these high frequencies.

\subsection{IRAM-30m telescope observations}
The observations of the \tco (2\rarrow1) and \cdo (2\rarrow1) were performed in February and March 2009 
using the HERA 3$\times$3 1mm receiver array at the IRAM-30m telescope located at Pico de Veleta (Granada, Spain). The observations were performed 
in OTF mode with the same OFF position as above. The spectrometer was the
VESPA autocorrelator,  configured to provide a spectral resolution of 40 kHz. Typical noise {\it rms} values are 0.2~K for both the \tco and 
\cdo lines.  The integrated intensity maps  are shown in Fig. \ref{fig_mapmon2}. 
In a different observing run, in July 2009, we observed  the H42$\alpha$ recombination line at 85.688 GHz. The spectrometer was the WILMA autocorrelator,
which provides a fixed spectral resolution of 2 MHz, $\sim$ 7 km s$^{-1}$ at this frequency. The noise {\it rms} value of these observations
is $\sim$0.05 K.

\begin{figure*}
\includegraphics[width=\linewidth]{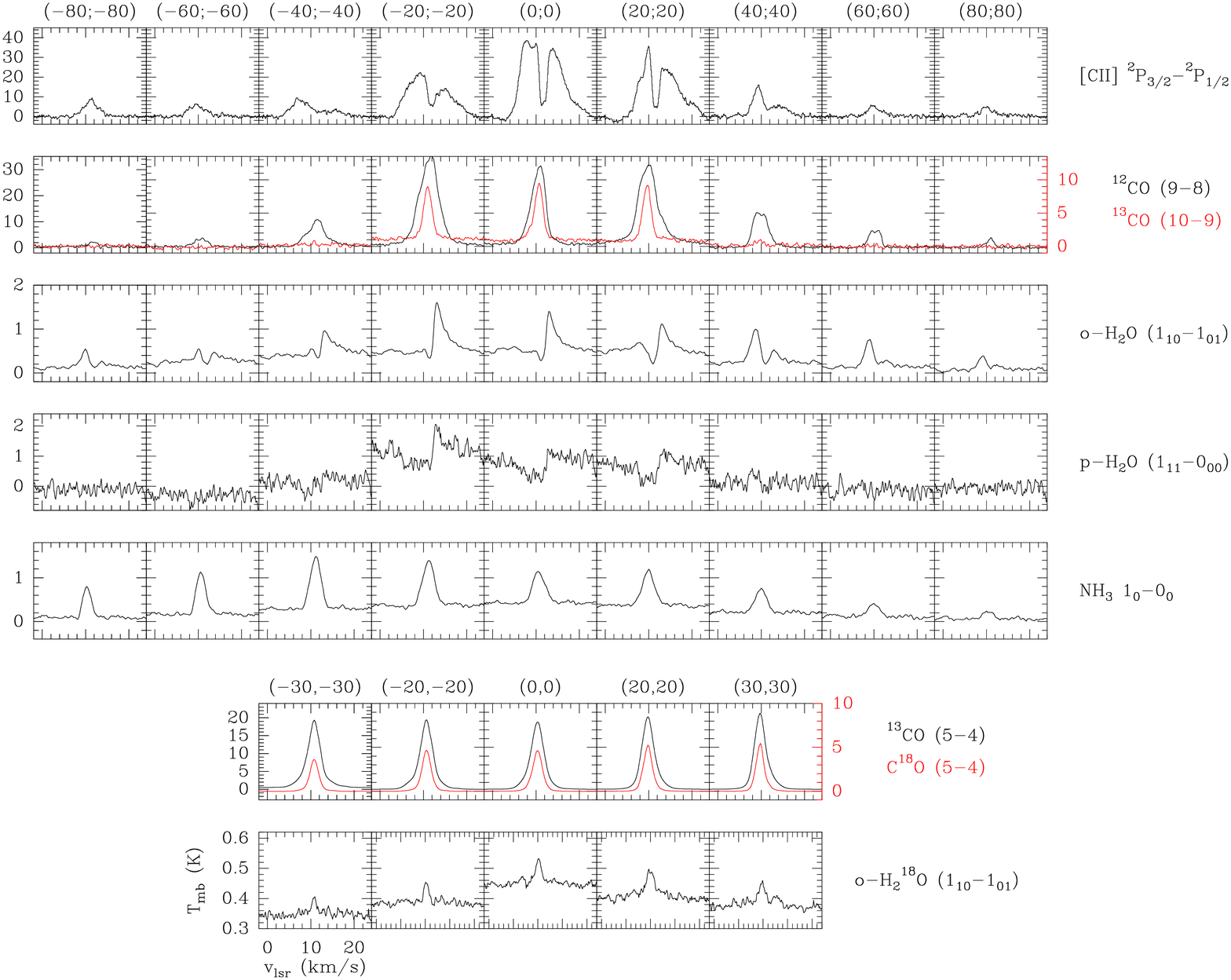}
\caption{ Raw spectra of the HIFI observations
along the cut indicated in  Fig.~\ref{fig_mapmon2}.  The continuum has not been subtracted so that the offset in the intensity scale represent the observed continuum. This needs  to be divided by two because of the double side band observations.  }
\label{fig_spectra}
\end{figure*}

\section{Results}
\label{sec_results}

\begin{figure*}
\includegraphics[width=\linewidth]{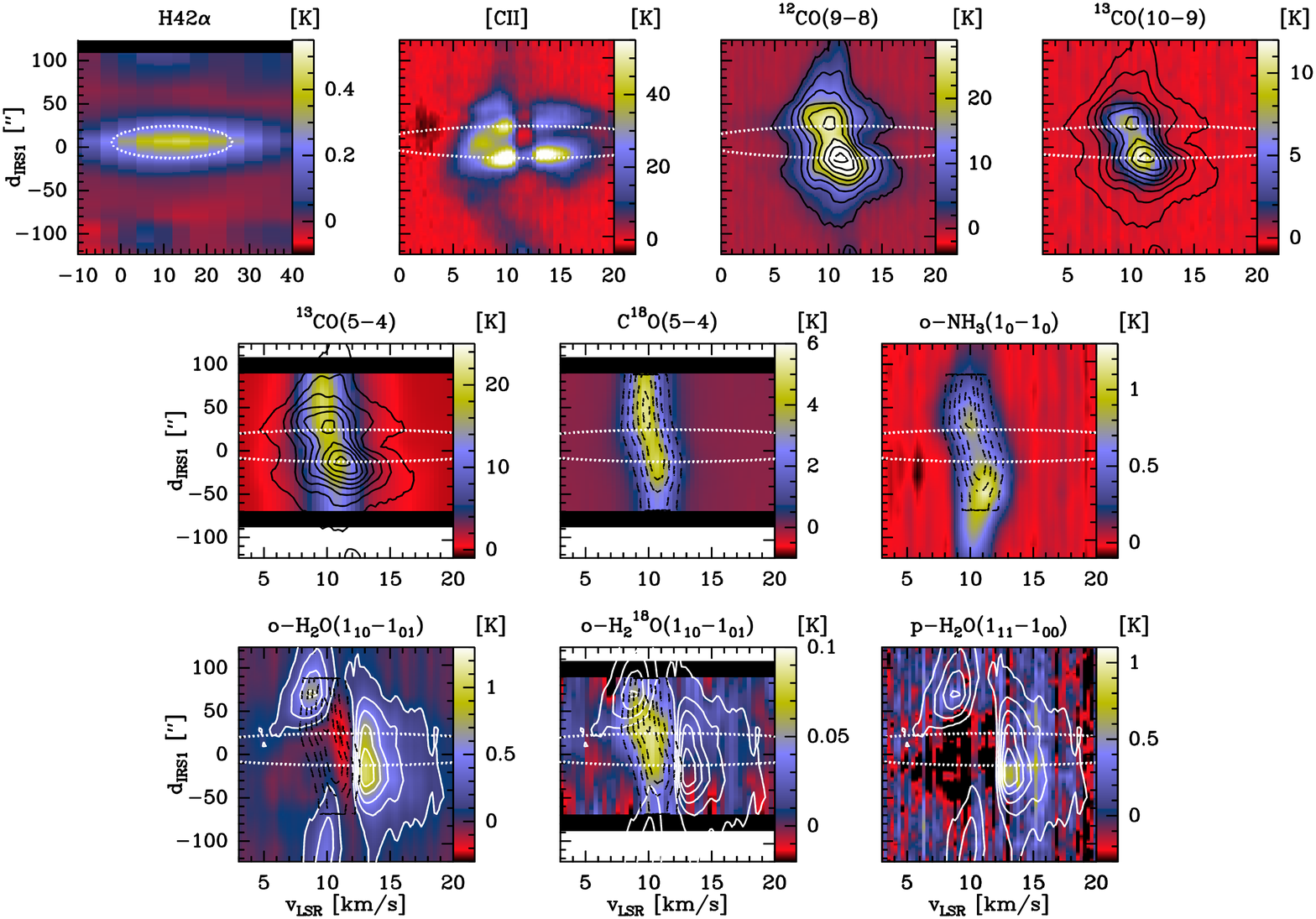}
\caption{Position-velocity diagrams along the observed strip (Fig.\,\ref{fig_mapmon2}) after continuum subtraction. The origin of the position axis is the IF and the negative direction extend SW of the origin.
The dotted white ellipse indicates the size of the \HII region. Black solid contours in the CO panels trace the intensity of \dco (9\rarrow8) emission 
from 2 to 37~K, in steps of 5\,K. Dashed black contours display the intensity of the \cdo (5\rarrow4) line from 1 to 6\,K in linear steps of 1\,K.
 The white solid contours represent the o-\wat 
emission from 0.1 to 0.9~K in steps of 0.2~K. 
}
\label{fig_pvs}
\end{figure*}

Figure \ref{fig_mapmon2} displays the integrated emission of the \tco and \cdo 2\rarrow1 lines in 
a 160\arcsec $\times$ 160\arcsec field centered at the IF. The emission of the  \cdo 2$\rightarrow$1 line presents   an arclike structure opened to the NW, with 
several peaks that correspond to regions of enhanced  gas density  as evidenced by other high-density tracers \citep{choi00, rizzo03}.  The shape of the integrated intensity of the \tco  
2$\rightarrow$1 line is similar,  although the emission is somewhat more extended and less sensitive to individual clumps.  Although it shows some symmetry, the  gas detected in \cdo shows an elongated structure extending northeast of the IF. The HIFI strips follows this structure so that individual clumps are   likely to influence the emission of far-IR lines in this direction rather than toward the SW, where the molecular emission is more compact. 

In the same figure, we present the integrated intensity of the H42$\alpha$ recombination line, which traces the \HII region. 
The large beam  ($\sim29\arcsec$  half-power beam width, HPBW) of the 30m telescope at this frequency  hinders the detection of the small-scale morphology
of  this UC  \HII region.
The peak intensity  of the H42$\alpha$ line is located $\approx3\arcsec$ northwest of the IF and the emission extends in a radius of  about 20\arcsec. 
Taking into account the large beam of the 30m telescope at this frequency and typical pointing errors of 1\arcsec-2\arcsec, we do not consider that
this shift is significant.

\subsection{Kinematics}

Figure \ref{fig_spectra} shows the  HIFI spectra  observed at different offsets along the the strip and Fig. \ref{fig_pvs}  displays their position-velocity (PV) diagrams. Several lines ([\CII], \dco 9\rarrow8, o-\wat and o-\amm) are  detected across the entire strip up to a distance of $\sim 100\arcsec$. 
The intensity and velocity structures show systematic variations along the strip, which can be summarized as follows:

\begin{enumerate}

   \item  The  [\CII] line is most intense and broader toward the UC \HII region, where it covers the velocity range from 0 to 20\,\kms. 
Outside the UC \HII region, the [\CII] emission remains detected but in a narrower 
velocity range, between 7 and 15 km s$^{-1}$. A probable origin of the high-velocity gas around the UC \HII region is a dense PDR layer, accelerated by the radiative pressure from the central source, where carbon is ionized. The extended and narrower emission is  likely related to the first quiescent layers of the molecular cloud or to layers in the external envelope.  
\item  The $^{12}$CO and o-\wat lines peak toward the UC \HII region  and high-velocity wings are detected in the direction of the ionized gas. Emission in the line wings is also 
detected at larger offsets toward the southwest (see offset [-40\arcsec,-40\arcsec]) in Figs.\,\ref{fig_spectra} and \ref{fig_pvs}). The origin of this highly accelerated gas is difficult to determine: 
at small offsets it may be associated to the same expanding layer that emits in [\CII], whereas the wings toward the southwest are more likely related with the
molecular outflow mapped by \citet{xu06}. Yet the outflow may also contribute to the broad emission toward the \HII region.
\item  The  \cdo and H$_2^{18}$O and \amm  lines have  relatively narrow line widths (between 2.5 and 3.1\,\kms) all along the strip.    The \tco, \cdo  and  H$_2^{18}$O lines peak a few arcsec northeast of the IF, whereas the \amm emission presents a dip toward the IF and peaks southwest. This extended emission is consistent with the overall structure of the cloud traced by the \tco and \cdo (2\rarrow1) lines, which present an elongated structure in the NE-SW direction. The dip in the \amm emission corresponds to a valley in the molecular bar traced by \cdo (2\rarrow1). 
\item An absorption feature  is detected in \wat at the velocity of the molecular cloud, $\approx 11.5$\,km s$^{-1}$, toward the UC \HII region, which likely originates in the same quiescent gas that is emitting for instance in \amm and \cdo.   The [\CII] spectra also show a self-absorption feature centered at $\approx 12$\,\kms but its origin is more doubtful.  In the other cases we do not detect the self-absorption feature.
\item  All CO  and \amm lines show a systematic shift of their peak velocities, which suggests that the expansion 
is not isotropic, and/or that a slow 
rotation \citep[${\rm v}_{\rm rot} \sim 0.5$\,\kms, ][]{loren77}  of the  cloud may be ongoing.

\end{enumerate}

  Summarizing,  these observations  suggest that the high-velocity wings
are coming from a gas that  is pushed away from the \HII region. This gas could be associated  either to the innermost layers of the expanding PDR, the outflow, 
or both.  On the other hand, the emission and the absorption at central velocities of \wat and H$_2^{18}$O, as well as the NH$_3$ and the low-J ($J_{\rm up} \leq 5$) CO  emission are more likely related 
to the interface of the PDR with the molecular cloud and to the bulk of the molecular cloud, respectively.  In the following sections, we model  the emission from CO and isotopologues, \wat and \amm with different degrees of approximation to derive the physical conditions at which they arise and derive an estimate of their abundances. The detailed analysis of the [\CII] line is postponed for a more detailed study that includes the [$^{13}$\CII] observations (Pilleri et al., in prep.).

\section{LVG modeling}

 In this section, we present  local radiative transfer calculations to determine the water and ammonia abundance toward Mon~R2, by assuming a uniform density and temperature layer. This approach is somewhat simplistic because  different phases of gas are mixed along the line of sight, but gives a first rough estimate of the water and ammonia abundance as a function of the physical conditions in the beam. 

For the sake of simplicity, we separated the line profiles into three velocity intervals, and analyzed each of them independently.  This allowed us to separate the optically thick and self-absorbed part of the line profiles from the optically thin wings.  The velocity intervals are defined as follows:

\begin{itemize}
\item $\Delta {\rm v}_b$ = [4.5-8.5]~\kms, 
\item $\Delta {\rm v}_c$ = [8.5-12.5]~\kms, 
\item $\Delta {\rm v}_r$ = [12.3-20.5]~\kms.
\end{itemize}

\begin{table*}
\caption{Mean intensities  ($<T_{\rm{mb}}>$) and LVG results for CO, \wat and \amm toward the IF after 
smoothing the observations to a 38\arcsec spatial resolution.}
\label{tab_observation-IF}
{
\begin{center}
\begin{tabular}{lcccc}
\toprule
& Units & 
$\Delta {\rm v}_b$  & 
$\Delta {\rm v}_c$ & 
$\Delta {\rm v}_r$  \\
& & [4.5-8.5] \kms & [8.5-12.5] \kms & [12.5-20.5] \kms \\         		    
\midrule
C$^{18}$O       (2$\rarrow$1)   & [K] &     0.58    &2.93 & 0.093      \\
C$^{18}$O       (5$\rarrow$4)   & [K] &   0.29    &   2.93   &  0.08        \\
$R^{18}_{5/2}$    		   & 		& 2		& 1	&     0.86                      \\
$N$(\cdo)/$\Delta{\rm v}$	& [cm$^{-2}$] & \ex{3}{14} & \ex{2}{15} & \ex{1}{14}\\
\midrule
o-H$_2^{18}$O    ($1_{10}\rarrow1_{01}$)   & [K] & $<0.01$  & 0.052 & $<0.01$    \\
R$_{H_2^{18}O/C^{18}O}$&                &-   &     0.018 & -     \\
$X$(o-H$_2^{18}$O)$\times 500$ 	&  &-  & $ [10^{-8}-10^{-7}]$ & - \\
\midrule
$^{12}$CO       (9$\rarrow$8)        & [K] & 5.11 & 24.86 & 2.83     \\
o-H$_2$O       ($1_{10}\rarrow1_{01}$)    & [K] &  0.08\tablefootmark{a}	& -0.13\tablefootmark{a}	&   0.59 \\
R$_{H_2O/^{12}CO}$&                     & - & - & 0.21        \\
$X$(o-H$_2O$) 	&  & -& -&$ [10^{-8}-10^{-6}]$  \\
\midrule
o-NH$_3$       ($1_{0}\rarrow0_{0}$)     & [K] &  $ < 0.01 $& 0.52 & $<0.04 $  \\
$R_{o-NH_3/C^{18}O}$&      &- & 0.18 &        -              \\
$X$(\amm)	&  &- & $ [10^{-9}-10^{-8}]$&- \\
\bottomrule
\tablefoottext{a}{Self-absorbed profile.}
\end{tabular}
\end{center}
}
\end{table*}

\begin{figure*}
\centering
\includegraphics[ width=0.49\linewidth]{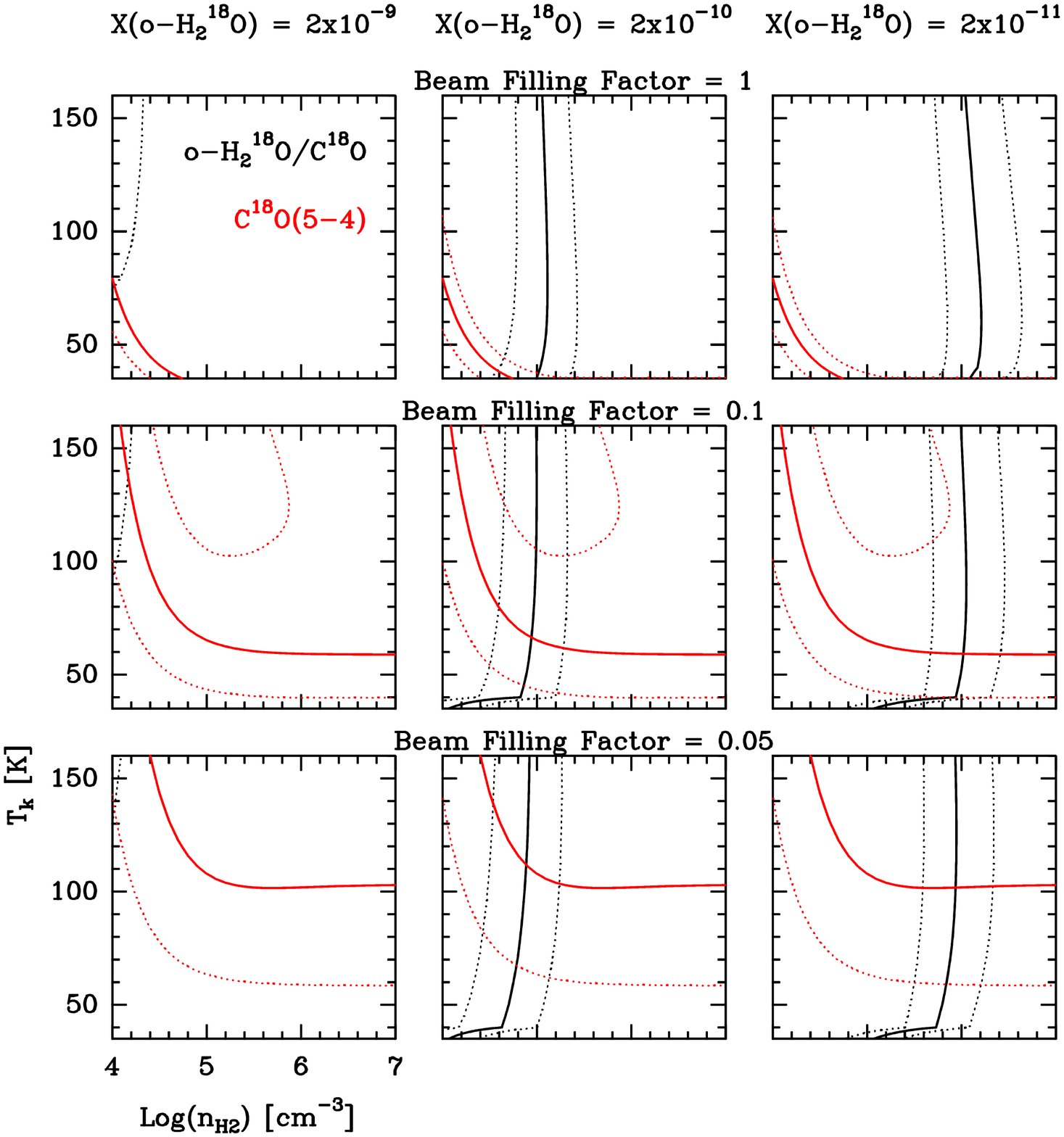}\hfill
\includegraphics[ width=0.49\linewidth]{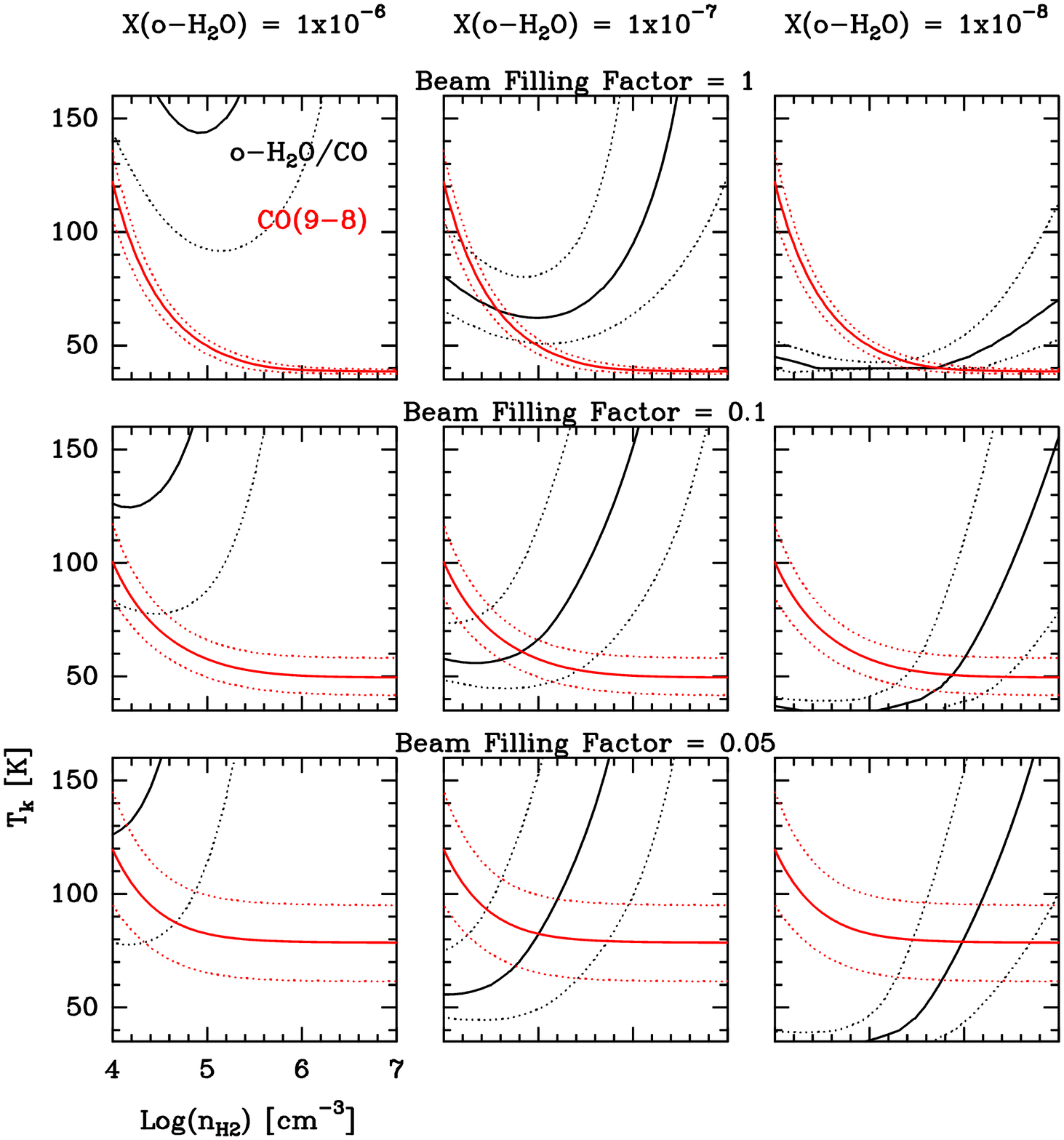}\\
\vspace{0.2cm}
\includegraphics[ width=0.49\linewidth]{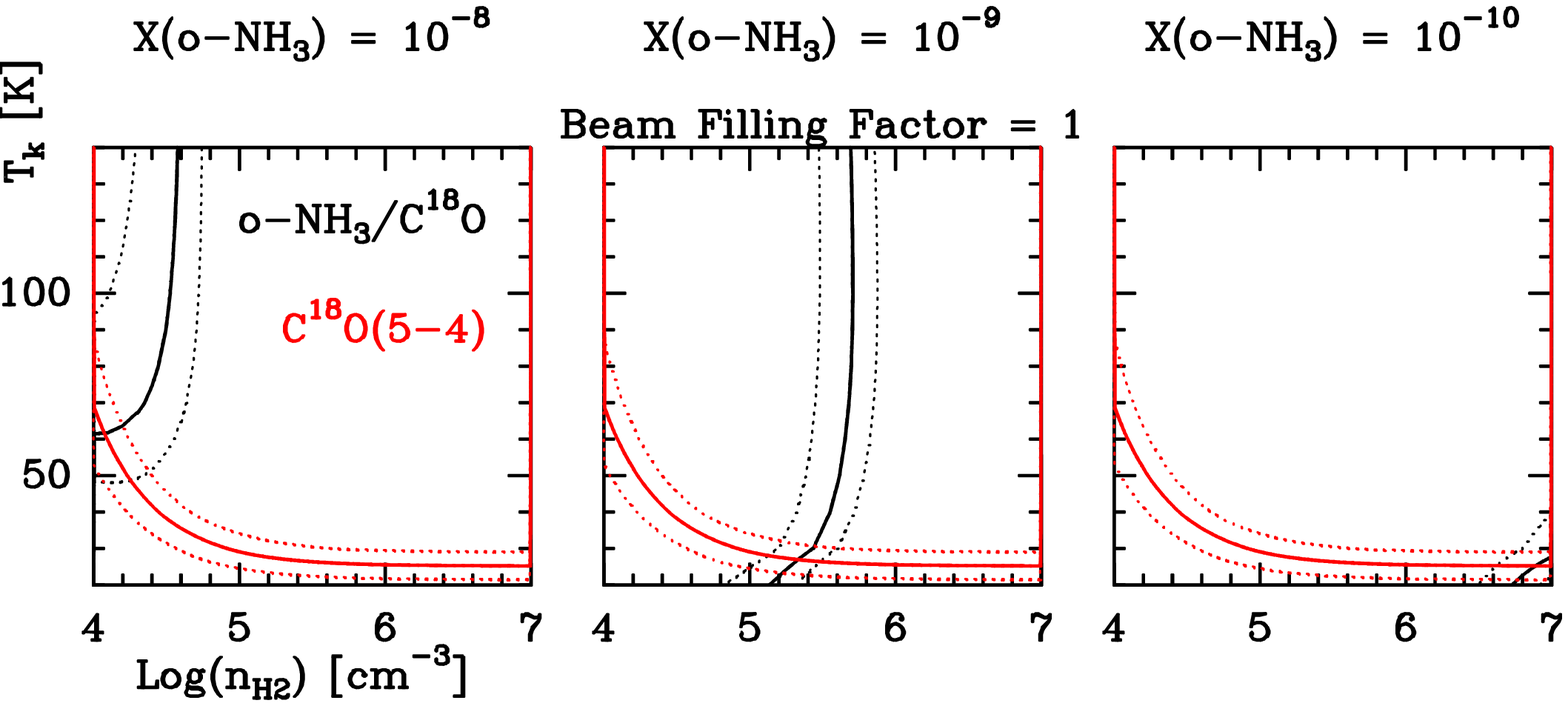}
      \caption{{\bf Upper left: } 1-D modeling of the intensity of 
          C$^{18}$O(5-4)    (red solid line) and of the ratio o-H$_2^{18}$O/CO(9-8) (black line) 
  in the velocity interval $\Delta {\rm v}_c$. 
    The intensities are calculated  as a function of the kinetic temperature and the local density, assuming different abundances of o-\wat and beam filling factors. The uncertainties (30\% in the absolute intensity and a factor of 2 for the line ratios)
 are represented with  dotted lines. {\bf Upper right:} The corresponding results for $^{12}$CO(9-8) (red solid line) and 
  o-\wat/\dco in the velocity interval $\Delta {\rm v}_r$.
    {\bf Bottom:} Results for the o-\amm observations in the velocity interval $\Delta {\rm v}_c$.  No solutions are found for a lower beam filling factor. }
         \label{fig_ratio_wat}
\end{figure*}

 For each of the velocity intervals, we used a large velocity gradients (LVG) code to fit the line intensities and estimate the column densities and abundances of  \wat \citep[MADEX, ][]{cernicharo12} and \amm \citep[RADEX, ][]{vandertak07}.
Table \ref{tab_observation-IF} shows the averaged main beam 
temperature measured per velocity interval 
($<T_b> = \frac{1}{\Delta {\rm v}}\int_{\Delta {\rm v}} T_{\rm{mb}}~d{\rm v}$) observed toward the IF.
 The gas column densities toward the IF were obtained by fitting the intensity of  the optically thin \cdo (2\rarrow1) and (5\rarrow4) lines as well as their ratio. 
Gas densities and kinetic  temperatures were varied in a reasonable range for this region 
($n_{\rm{H}_2}=10^4-10^7$\,\cmc, $T_{\rm k}=40-500$ K) and
the grid was repeated for different values of the beam filling factor and abundance.  

\subsection{H$_2^{18}$O emission}
 We modeled the intensity of the optically thin line  o-H$_2^{18}$O (1$_{1,0}$$\rightarrow$0$_{0,1}$) to obtain a first estimate of the water  abundance  in $\Delta{\rm v}_c$.
We fit the intensities of the C$^{18}$O (5$\rightarrow$4) and o-H$_2^{18}$O (1$_{1,0}$$\rightarrow$0$_{0,1}$) lines and their ratio, assuming a standard isotopic ratio $^{16}$O/$^{18}$O=500 and a \dco abundance relative to \hd  of \mbox{$X({\rm ^{12}CO})= \ex{1}{-4}$.}  
We used the new collisional rate coefficients by \citet[][]{daniell11} for o-\wat and \citet{yang10} for CO. We also assumed an  \hd ortho-to-para ratio (OTP) of 3.  Since the collisional coefficients of \wat with ortho- and para-\hd species are asymmetric, the results depend on the assumed OTP ratio.  In particular, assuming a lower OTP ratio would result in the need of higher densities to excite \wat. 

 Figure \ref{fig_ratio_wat} shows the results of the LVG modeling. 
The  intensities of the lines and their ratio can be reproduced reasonably well assuming a beam filling factor  ($\eta_{\rm ff}$) of 0.1 or lower. 
Several solutions can be found with this $\eta_{\rm ff}$, corresponding to different  density-abundance pairs. The line intensities can be fitted with several combination of these two parameters, which reflects a 
 degeneracy between the gas density and the water abundance. Considering the range of densities that are expected to be found in Mon~R2, the solutions correspond to a water abundance between $10^{-7}$ and $10^{-8}$ relative to H$_2$. 
The highest value is obtained if 
the emission arises from relatively low-density gas ($n_{\rm H_{2}} \sim$ a few $10^4$\,\cmc), whereas the lowest abundance corresponds to gas densities of $10^{6}$\,\cmc.

\subsection{H$_2$O emission}
Because the \wat lines are heavily self-absorbed at central velocities, we concentrated our LVG analysis on the emission in the red wings, in the velocity interval $\Delta v_r$.  
Emission from \dco (9-8) and o-H$_2$O (1$_{1,0}$$\rightarrow$0$_{0,1}$) is expected to arise in similar physical conditions, because both lines need relatively 
high temperatures and densities to be excited.
  We modeled the intensity of the o-\wat emission and the ratio $R_{H_2O/^{12}CO}=I($o-H$_2$O 1$_{1,0}$$\rightarrow$0$_{0,1}$)/$I(^{12}$CO 9$\rightarrow$8)  to derive the water abundance in the line wings.

Assuming a beam filling factor $\eta_{\rm{ff}}= 1$, which would imply an angular extension of 
the region emitting in the red wings of $\sim$38\arcsec, we only found solutions for  o-\wat abundances between 10$^{-8}$ and 10$^{-7}$ relative to \hd. 
The highest water abundance corresponds to a molecular hydrogen density of $\sim5\times 10^5$ cm$^{-3}$ and gas kinetic temperature of about 70 K.  For an
abundance $\sim$10$^{-8}$, densities of several $\sim$10$^5$\,\cmc and  temperatures $\sim$50 K are needed to reproduce the measured line intensities. Since both set of values
are within the range of physical conditions that we can  accept for this region, we cannot  easily distinguish between them. 

To assume a beam filling factor of 1, however, is  a limiting case, since the high-velocity wings are concentrated in the PDR around the UC \HII region. 
  A beam filling factor $\eta_{\rm{ff}}$=0.1   would 
correspond to the case of   an emission ring extending only in the plane of the sky  with a thickness 
of $\sim$2\arcsec\, and a diameter of 20\arcsec, which is consistent  with the size of the PDR traced by  PAH emission around this region \citep{berne09a}.
 Assuming these lower values of  $\eta_{\rm{ff}}$, we found  a water abundance $X$(o-H$_2$O) in the range [$10^{-8} - 10^{-6}$] for reasonable physical conditions 
in this region ($n_{\rm{H}_2}$  of several $10^4 - 10^6$cm$^{-3}$ and $T_{\rm k}$ between 50 and 100 K).
These values of molecular hydrogen density are consistent with the  physical conditions derived by  \citet{berne09a} 
on the basis of the purely rotational lines of \hd that trace the PDR.  This supports the scenario of an expanding PDR in which the highest velocities 
are associated  to the molecular gas closer to the UC \HII regions, i.e., the warm HI/H$_2$ layer traced by the PAHs and the H$_2$ rotational lines. 
However, the possible contribution of an outflow component cannot be discarded, especially to the southwest, where the outflow detected by \citet{xu06} has its maximum
emission.

\subsection{\amm}

We used the same approach to estimate the o-\amm abundance in $\Delta{\rm v}_c$. The profiles
of the o-\amm lines are not self-absorbed, and can be used to obtain a direct  estimate of the column density of this species for all velocity intervals. 
In contrast to water,
the $R_{NH_3/^{12}CO} = I($NH$_3 1_0\rightarrow0_0)/I(^{12}$CO 9$\rightarrow$8) ratio decreases at higher velocities, corroborating our interpretation that the NH$_3$ emission does not 
come from the innermost part of the PDR, but from the bulk of the molecular cloud. 

  We have fitted the intensity of the o-\amm line and its ratio to the \cdo (5\rarrow4) line using the collisional  rate coefficients from \citet{danby88}.  The results for $\Delta {\rm v}_c$ are reported in Fig. \ref{fig_ratio_wat}. 
Assuming a beam filling factor of $\sim$1, consistent with the large extension
observed with HIFI, our data are  reasonably well reproduced with a density of $\sim10^4-10^5$ cm$^{-3}$, a gas kinetic temperature of $\sim$50K and
an ammonia abundance $\sim10^{-8} - 10^{-9}$. The opacity of the o-\amm line are of the order of $\tau \sim 10$, so the lines are very optically thick.  Lower values of the \amm abundances are associated to higher densities that have no evidence in MonR2. %

\section{A simple spherical model of MonR2}

  In this section, we test a simple geometrical model of the region to reproduce the observed line profiles. The model assumes an expanding spherical structure composed of concentric layers with given physical conditions (gas temperature and local density),  kinematical information
(turbulent and expansion velocities), and molecular abundances. 
 We  note in advance that with such a simple, spherically symmetric model is impossible to account for all asymmetries in the velocity structure of the observations, which may reflect large-scale structures such as the outflow, small-scale inhomogeneities (clumpiness) and other kinematical effects such as rotation. Yet, we propose it as a useful step toward a better understanding of this region.

In a first step, we  fit  the CO observations to fine-tune the physical structure of the molecular cloud and its associated PDR. This structure is then used to fit the  \wat observations by varying their abundance profiles across the cloud. To keep the model as simple as possible, we assume a double step-function abundance profile. For  \amm,  our observations consist of a single, optically thick line that does not provide a reliable probe of the detailed spatial distribution of the o-\amm abundance, and we therefore exclude this molecule from this toy model.

\subsection{The physical structure}
The physical structure (in terms of the length of each layer, their density and temperature) is derived based on previous observational data and modeling \citep{berne09a, fuente10}. 
 A first estimate of the gas temperature was obtained using an updated version (1.4.3) of the  Meudon PDR code\footnote{\url{http://pdr.obspm.fr/PDRcode.html}} \citep[][ cf. also Sect.\,\ref{subsec_gasphase} and Table~\ref{table_meudoninput}]{lepetit06,goicoechea07,gonzalez08,lebourlot11}.

 The density structure in Mon~R2 can be assumed to be similar to other PDRs associated to star-forming regions, such as NGC 7023 and the Orion Bar \citep[in prep.]{joblin11}.  The first 
layer ($\sim$ 1-2 $A_{\rm{V}}$) of the PDR is where most of the [\CII] and PAH emission originates from \citep[][]{habart03, joblin10,pilleri11}. This is generally followed by a 
high-density  layer (or "filament") with $n_{H_2}\gtrsim 10^5$\,\cmc,  which can be several magnitudes thick and is responsible for the emission in the high-J rotational 
lines of CO. Finally, the bulk of the molecular cloud is usually at a lower density, and accounts for the emission of molecular tracers that are easily photo-dissociated. In our model, the outermost layer (3 mag thick) is heated from the outside, in agreement with the large large-scale PAH emission detected at 8$\mu$m, which testifies to the presence of an external source of heating, especially to the north of the \HII region \citep{ginard11}.

 In our spherical model, the central core of the `onion' is represented by an \HII region with a very low density that is free of molecular gas. 
The \HII region is surrounded by a warm and relatively low-density expanding layer ($L_{PDR}$). For this layer we adopted the density derived by \citet{berne09a} from the H$_2$ rotational
lines. The next layer is the high-density layer, $L_{HD}$ ($n_{\rm{H}_2}$$\sim\ex{{3}}{6}$~\cmc). $L_{HD}$  extends for a few 0.001~pc  ($\sim$ 10 mag) and is required  
to explain the emission  of large dipole molecular tracers such as CS, c-C$_3$H$_2$ and HCO$^+$ \citep{ginard11}.
Finally, everything is surrounded by a lower-density evelope, $L_{env}$ \citep{fuente10}. The inner radius of the PDR, $r_{\HII}$= 0.08~pc ($\sim20\arcsec$), is determined by the angular 
size of the \HII region, and the outer radius of the envelope, $r_{out}$=0.34~pc, is fixed on basis of the extension of the $^{13}$CO 2\rarrow1 and C$^{18}$O 2\rarrow1 line emissions  
(Fig.~\ref{fig_mapmon2}). The values of $r_{PDR}$ and $r_{HD}$ are tuned to fit the CO (and isotopologues) lines.
Summarizing:
$$
n_{\rm{H_2}} (r) = \left\{ \begin{array}{cll}
	\ex{2}{5}~{\rm cm}^{-3} &\mbox{   if $r_{\HII} < r < r_{PDR}$} & \rightarrow L_{PDR}\\
	\ex{3}{6}~{\rm cm}^{-3} &\mbox{   if $r_{PDR}   < r < r_{HD}$}& \rightarrow L_{HD}\\
	\ex{5}{4}~{\rm cm}^{-3} &\mbox{   if $r_{HD}   < r < r_{out}$}& \rightarrow L_{env}\\
       \end{array} \right. 
$$

For the kinematics, we adopted the expansion velocity law
\begin{equation}
\label{eq_velocity}
V_{exp} (r)= V_{out} \times (r_{out}/r),
\end{equation}
\noindent in which $r$ is the radial distance from the center of the sphere and $V_{out}$ is the expansion velocity at $r_{out}$. 
 This law mimics the expansion velocity profiles commonly used to model \HII regions, with the surrounding envelope expanding at a lower velocity compared to 
the PDR and the \HII region \citep{lebron01}. 

Finally, we included a continuum source in our model to reproduce the  fluxes measured with \herschel. 
The continuum fluxes toward the IF are well reproduced  by a modified black body at the temperature of 500~K with an opacity 
of 0.1 at 50~\um,  an exponent for the opacity law of 1 and a radius of 0.03\,pc. This enables one to adequately fit the continuum levels toward the IF, but underestimates it at larger offsets. The presence of other continuum sources in the background might help to reproduce the continuum levels at large distance from the IF, but these are not included in our model for simplicity.

\label{sect_ model}
\subsection{CO lines}
\label{sect_co}

We used our non-local radiative transfer code \citep{cernicharo06} to fit the line profiles
at various offsets along the HIFI stripes. 
 We adopted a standard $^{12}$CO abundance relative to H$_2$ of 10$^{-4}$ 
and the isotopic abundance ratios of $^{12}$C/$^{13}$C=50 and $^{16}$O/$^{18}$O=500.

 Assuming the
density structure described above and the gas kinetic temperature derived with the  Meudon PDR code, we fine-tuned the values of $r_{PDR}$ and $r_{HD}$ and
the kinematical parameters to best reproduce the CO observations. The integrated intensities are well-fitted with $r_{PDR}$ and $r_{HD}$ being 0.082 pc and 0.083 pc, respectively. 
 To improve the fit of the CO lines, the mean kinetic temperature of the outermost 3 mag of the cloud is set to 50\,K, which corresponds to an external radiation field of $G_0^{ext} \sim$ a few 100. This value is consistent with an extended PDR at a  distance of $\sim 100\arcsec$, which is illuminated by the UV field produced by IRS1. 

\begin{figure}
\centering
\includegraphics[trim = 0cm 0cm 0cm 0cm, width=0.8\linewidth]{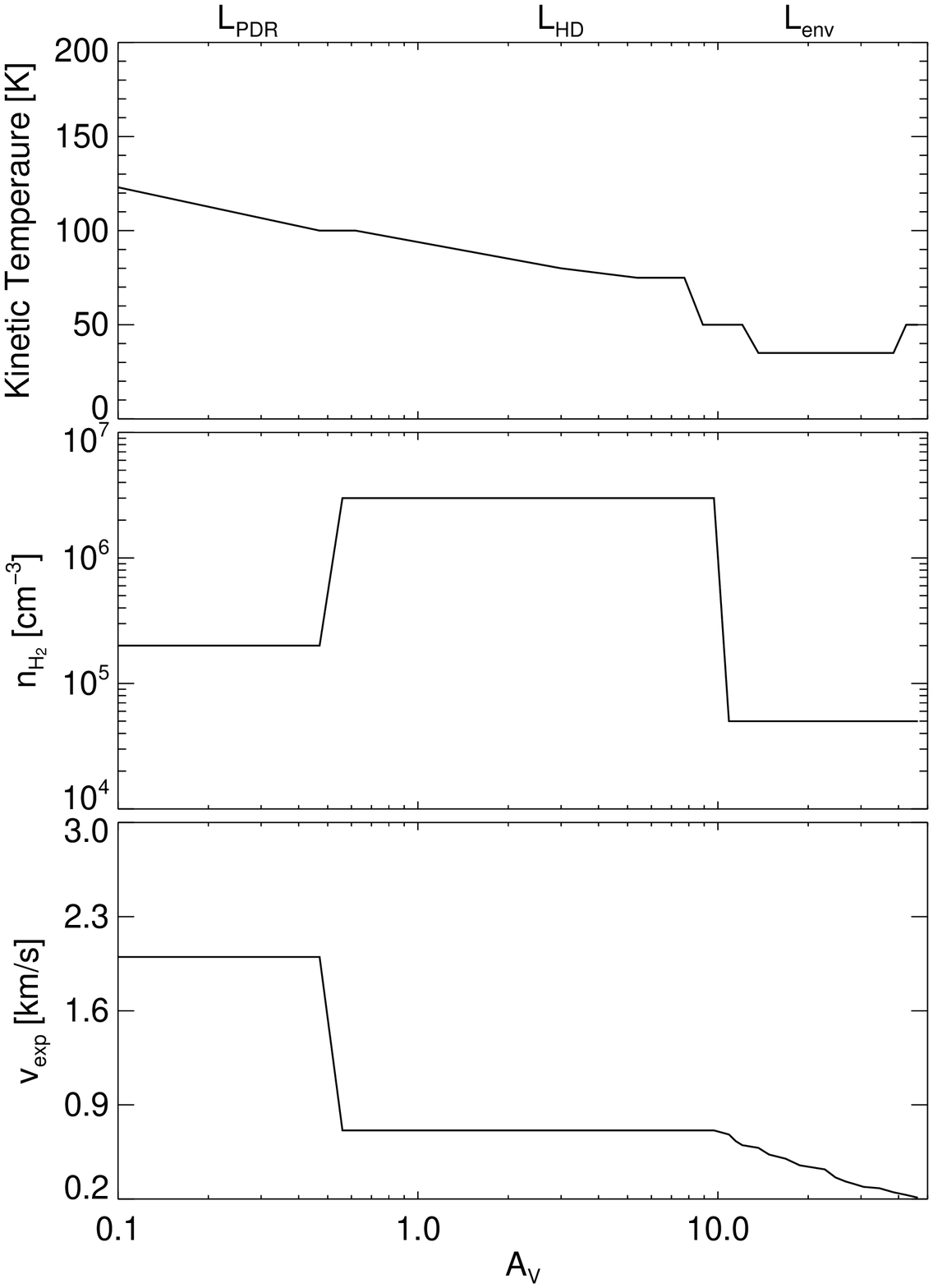}
      \caption{Summary of the physical and velocity structure used in the  non-local radiative transfer model. Values for $A_{\rm V} \le 0.1$ are constant. }
         \label{fig_physcond}
\end{figure}

The best fit to the line profiles was obtained using  the velocity law of Eq.\,\ref{eq_velocity} with $V_{out}=0.125~{\rm km\,s}^{-1}$. 
We have also varied the turbulent velocities to improve the fit of the wings of the \dco lines. The best solution is found with a turbulent velocity of 
1\,\kms  in $L_{HD}$ and 1.5\,\kms in L$_{env}$. To reproduce the high-velocity wings, we fixed the expansion and turbulent velocity in $L_{PDR}$ to  2\,\kms and 5\,\kms, respectively. 
We stress that because the model has so many parameters, other solutions may be found that reproduce 
the CO emission well, especially in the wings. Higher resolution observations will be very useful to improve our knowledge of the structure of this region. Figure \ref{fig_physcond} shows a summary of the physical structure and the velocity profile across the PDR. 

\begin{figure*}
\centering
\includegraphics[ angle = 270, width=1\linewidth]{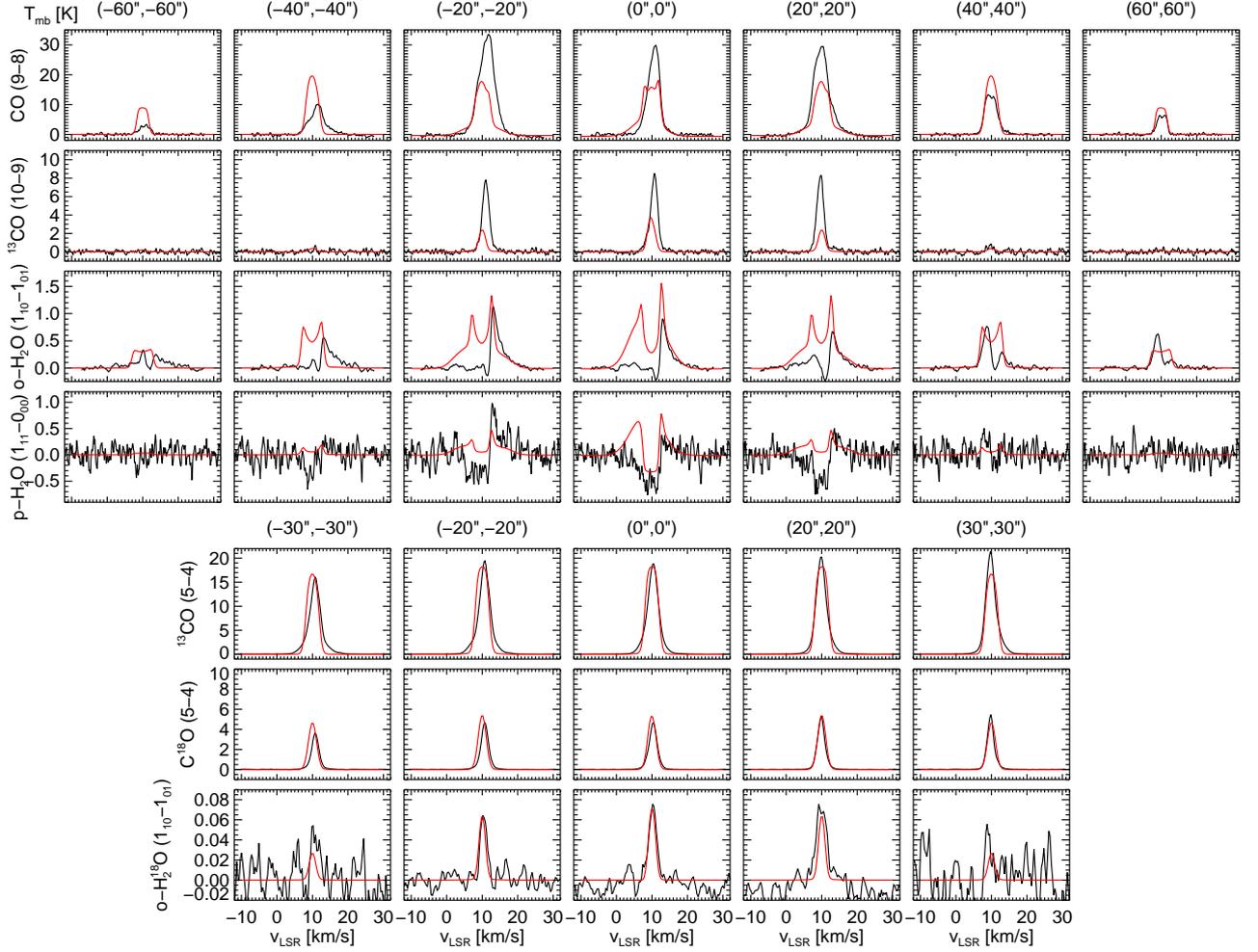}
      \caption{Comparison of the continuum-subtracted HIFI spectra (black) and the results of the non-local radiative transfer modeling (red) assuming a spherical symmetry for Mon~R2.  
 }
         \label{fig_madex}
\end{figure*}

Figure \ref{fig_madex} displays the model results and the comparison with the observations.   Although the overall quality of the low-J lines is very good, there are some discrepancies. The model falls short of the high-J ($J_{\rm up} >9$) CO lines at offsets [-20\arcsec, -20\arcsec] to [+20\arcsec,+20\arcsec]. 
This is not unusual in PDR, because current PDR models fail to reproduce the intensity of the high-J lines of CO. There are various effects that can be invoked to explain this missing intensity, i.e. clumpiness and an erroneous description of the microphysics of PDRs, which may influence the heating and the chemistry of PDRs \citep[][in prep]{joblin11}.
The presence of a high-velocity outflow, or the inclusion of rotation may all be possible ways to improve the fit. 
However, this accurate level of modeling is beyond the scope of this paper.

\subsection{H$_2^{18}$O and H$_2$O}
 We used the physical and kinematical structure derived above to constrain the abundance of water in each  layer of the cloud. 
 We assumed an abundance profile defined as a double step-function and the standard ortho-to-para ratio of 3.  We varied  the abundances in the three regions to reproduce the line profiles of all  \wat lines. 
 
The emission of H$_2^{18}$O is optically thin and does not present significant emission in the wings, and therefore its intensity is influenced mainly by the low-velocity layers $L_{HD}$ and $L_{env}$.  
We were able to reproduce the intensity and profile of this transition at all offsets by assuming an abundance $X($o-\wat)$ = \ex{3}{-8}$ in $L_{HD}$ and  \ex{1}{-8} in $L_{env}$. These values are consistent with the results of the local LVG modeling assuming a beam filling factor $\lesssim 0.1$ and with the previous estimate reported in \citet{fuente10}.  With the assumed physical structure, small variations of the water abundances (more than a factor of 2) from these values do not allow one to fit the observed intensities and spatial distribution. However, the main source of uncertainty is the assumed values for $n_{\rm H}$ and $T_{\rm kin}$. Considering the range of physical conditions that is observed in Mon~R2, we can estimate our final error bars in the 
water abundances to be of one order of magnitude.

With the same velocity profile, we tried to reproduce both the {\it ortho-} and {\it para-}\wat lines at all offsets (See Fig.\,\ref{fig_madex}). However, both lines are very optically thick and  variations in the morphology and kinematics of the cloud can strongly modify the line profiles. 
Therefore, we concentrated our efforts on the modeling of the red wings only, which are optically thin and therefore less sensible to small variations of the detailed velocity structure.
To reproduce the line wings, we varied the abundance of \wat in the layer $L_{PDR}$. Our best fit is obtained with an abundance of  $X($o-H$_2$O$) = \ex{1}{-7}$ in this layer, consistent with the results of the LVG modeling assuming a  density of a few $10^{4}$ and high kinetic temperature (between 50 and 150 K). 
 Similarly to H$_2^{18}$O, the main source of uncertainty for \wat are the model parameters. However, the physical conditions in this layer are relatively well constrained  \citep{berne09a} and the corresponding water abundances are in the range [$10^{-7}-10^{-6}$] relative to \hd.    
As mentioned above, the complete velocity profile  of the main isotopologues is very dependent on the assumed kinematics and abundance. Therefore, it is difficult to obtain a very good fit of all spectra with a very simple model like ours. 
There are a few discrepancies between the observed \wat profiles and the model. The observations are not symmetric between positive and negative offsets, and there is often an excess emission in the blue wings.  These differences  are most likely due to the unrealistic assumption that our cloud is perfectly spherical and symmetric 
under any point of view. The PV diagram of [\CII] clearly shows that the kinematical pattern is not symmetric. The wings are wider at red velocities, probably because of the cometary shape of the nebula. 
In our model, we have used the red wings as the basis for our expansion pattern which is very likely incorrect
for the blue part.

Summarizing, we derived a relatively constant (\ex{1-3}{-8}) abundance of water in the high-density region and in the envelope, and a larger abundance (\ex{1}{7}) in the high-velocity gas. With these abundances, we obtained very good fits to all three water lines at the same time. 
With the assumed physical structure, the abundances are relatively well constrained (within a factor of 2), since small variations of the water abundance have a strong impact on all the lines. Larger variations from these values do not enable one to reproduce all th lines at the same time. However, since the density and velocity structure are not very well constrained and because of the degeneracy between density and abundance, we speculate that a reasonable uncertainty on the water abundance is one order of magnitude. 

\section{Discussion}
\label{sec_discussion}

\subsection{\wat abundances in MonR2 and the ISM}

 The results of the previous sections on the water abundance can be summarized as follows. In the gas at low velocities, the abundance of water is $\sim$ a few $10^{-8}$ relative to \hd.
The spatial distribution and the radiative transfer analysis of the H$_2^{18}$O line suggests that the low-velocity water emission arises in a region of enhanced density and relatively small volume centered toward the IF. This  corresponds to  the physical conditions of the high-density  PDR proposed by \citet{rizzo03}. 

In the high-velocity gas, the abundance seems to be about an order of magnitude higher and the origin of the wing emission is more doubtful. 
  In our simplistic `onion' model of Mon~R2,  both the red emission and blue absorption are due to the expanding gas associated with the PDR surrounding the \HII region  \citep{rizzo05, fuente10}. 
  In this scenario, the emission in the red wing of the o-H$_2$O line would be related to the PDR around the UC \HII region. 
 This is consistent with the high-velocity wings observed in 
typical PDR tracers such as c-C$_3$H$_2$ and the small linewidth of shock tracers such as  SiO \citep{rizzo03}. 
However, our simple model cannot reproduce the  detailed spectral profiles of the \wat lines. In particular, our model overestimates the blue-shifted emission  
at every position in the strip and it cannot reproduce the red wings at the offsets (-40\arcsec,-40\arcsec) and (-60\arcsec,-60\arcsec) (see Fig. 6). An asymmetric expanding PDR, and the contribution of the
molecular outflow to the water line profiles are very likely the cause of this discrepancy.

 The values derived in this work for the water abundances are similar to those commonly found in massive star-forming regions, and somewhat lower than those usually found in outflows and hot cores. 
 In early-stage massive star-forming regions such as W3 IRS5, the abundance of water has been shown to vary between $10^{-9}$ in the outer and cold envelope and $10^{-4}$ in the inner hot cores \citep{chavarria10}. Concerning lower mass star-forming regions, \citet{lefloch10}  studied the water emission in the bipolar outflow of L1157, determining a water abundance varying between few $10^{-7}$ in the warm and dense ($n>10^5$\,\cmc, $T_{\rm K}\sim100$~K) extended part of the outflow  and few $10^{-5}$ in the high-velocity hot component ($T_K>100$~K) that arises from the bow shock. 
 \citet{kristensen10, kristensen11} derived similar abundances in the early stages of low-mass star-forming regions. 
 
  In agreement with previous  estimates of the water abundance toward Orion KL based on the IRAM-30m telescope \citep{cernicharo94} and ISO data \citep{cernicharo06b}, most recent observations with \herschel \citep{melnick10} have derived a  water abundance of $\sim 10^{-5}$ in the 
  outflow, much higher than that derived in this work.
Using spatially resolved PACS observations, \citet{habart10} derived an upper limit to the water abundance  in the Orion Bar of a few $10^{-7}$. This value is very consistent with our results. 
 Still unpublished data from the WADI consortium show that water is detected also in PDRs not associated with outflows, such as NGC 7023, Ced 201 and the Horsehead nebula (Teyssier et al., in prep.). The water abundance derived in the Horsehead nebula seems to be much lower than that in MonR2, most likely because the dust temperature is very low ($\sim 30$\,K), and sticking on grains is more efficient. 
Interestingly, in the PDR associated to the proto-planetary disk TW Hydra, \citet{hogerheijde11} 
derived an upper limit to the abundance of water of  \ex{0.5-2}{-7}, which is consistent with the values detected in the PDR of Mon~R2.

Summarizing, the abundances measured in  Mon~R2 are lower than those obtained in shocked regions,
similar to those observed in the outer layers of pre-stellar cores and the envelopes
 of young stellar objects, and consistent with the upper limit derived toward the Orion Bar.  

\subsection{Comparison with gas-phase PDR chemical models}
\label{subsec_gasphase}

It is interesting to compare our observational results with the predictions of the Meudon PDR chemical model to understand where gas-phase chemistry is sufficient to explain the observed water abundances, and where gas-grain chemistry needs to be taken into account. The input parameters to the code are reported in Table \ref{table_meudoninput}.  
We have tested the impact of using different extinction curves (a standard galactic and the Orion Bar's) as well as  varying the cosmic ray ionization rate. We did not find significant differences in the predicted water abundances  in the gas phase.

\begin{table}[btph]
   \centering
   \begin{tabular}{lll} 
 \toprule
 	Parameter & & Value \\
	\midrule
	      $ G_0$    & radiation field intensity &  \ex{5}{5} \\
	   $ G_0^{ext}$    & external radiation field intensity & 100 \\
    $A_{\rm V}$& total cloud depth &50 mag	\\
    &   extinction curve & standard galactic\\
    $R_V$	& $A_{\rm V}$/$E_{\rm B-V}$ & 3.1 \\
     $\zeta$	 & cosmic ray ionization rate & \ex{5}{-17}~s$^{-1}$	\\
    $a_{min}$  & dust minimum radius& \ex{3}{-7} cm	\\
    $a_{max}$ & dust maximum radius & \ex{3}{-5} cm	\\
    $\alpha$ & MRN dust size distribution index   &3.5\\
    He/H & Helium abundance & 0.1  \\
    O/H & Oxygen abundance &  \ex{3.2}{-4}  \\
    C/H & Carbon abundance &  \ex{1.3}{-4}  \\
    N/H & Nitrogen abundance & \ex{7.5}{-5}  \\
    S/H & Sulfur abundance & \ex{1.8}{-5}  \\
     \bottomrule
   \end{tabular}
   \caption{Input parameters for the Meudon PDR code}
   \label{table_meudoninput}
\end{table}

\begin{figure}
\centering
\includegraphics[ width=1\linewidth, , trim = 1cm 1cm 1cm 1cm]{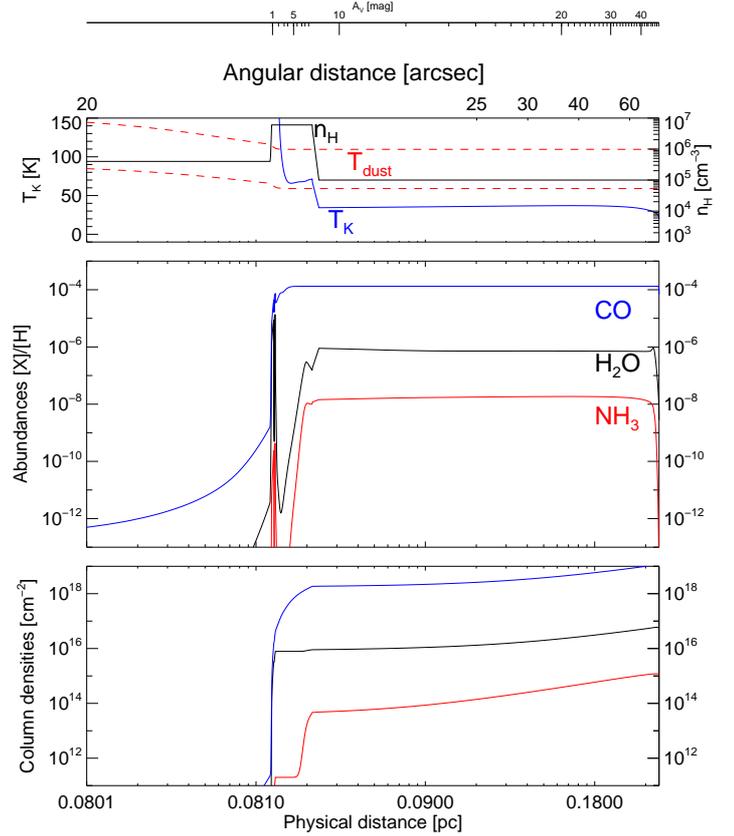}
      \caption{Results of the modeling of MonR2 with the Meudon PDR code. The angular and physical distances refer to the projected distances from the IF in the plane of the sky.  The two curves for the dust temperature are for grains of radius \ex{3.6}{-7} and \ex{3.6}{-5}\,cm.
 }
         \label{pdrmeudon}
\end{figure}

The  water abundance predicted by the Meudon PDR code (middle panel in Fig.~\ref{pdrmeudon}) shows significant variations with $A_V$. Following a very low abundance in the most exposed 
layers ($A_{\rm V} \lesssim 1$), the code predicts a water abundance  of  X[H$_2$O]$\sim 10^{-7}-10^{-5}$ in  a compact layer of about 1 mag.  Subsequently, the water abundance strongly dips 
 to $10^{-11}$ in the high-density layer at about $\sim$5~mag, and increases again 
and reaches its maximum value of $\sim10^{-6}$ for $A_{\rm V} \gtrsim 10$.

 Our results show that the o-\wat abundance is $\sim 10^{-7}$ in the first layers of the PDR ($L_{PDR}$). This is consistent with 
the value predicted by the Meudon code under the assumption that the outflow has only a minor contribution to the wing emission at central positions.
In this layer, water is formed in gas phase via
\begin{align*}
\rm{OH} + H_2	\rarrow	& \hspace{0.2cm}\rm{H}_2\rm{O}+\rm{H}	.\\
\end{align*}
This reaction has a  high activation barrier ($\sim4000$\,K) and is significant only at $T_{\rm{K}}\gtrsim 250$~K.  The internal energy available in vibrationally excited \hd can help to 
overcome this barrier \citep{agundez10}. Photodissociation is the main destruction mechanism.
 
 In the rest of the cloud, the gas-phase predictions and the observational results do not agree.
 In the high-density region, the predicted water abundance is lower than $10^{-10}$, whereas the H$_2^{18}$O measurements point to a relatively high abundance of  $\sim10^{-8}$. An additional
 water supply mechanism, such as desorption (photo-desorption, mechanical sputtering, thermal desorption) from the grain mantles, is required. In contrast,  
 the measured \wat abundance  in the envelope is at least two orders of magnitude lower than  the abundances derived by the chemical model. The freeze-out of water onto grain mantles is
 very likely an efficient gas-phase water destruction mechanism in this colder region. Although it is difficult to compare these results 
 without a precise profile for the water abundance, these results clearly show that gas-phase chemistry is not sufficient to account for the water emission in this PDR.

We have also compared our results with the predictions from the PDR model of  \citet{hollenbach09}, which includes  freeze-out and photo-desorption processes but does not consider  turbulence. 
Our observational value of the water abundance in the envelope (\ex{1}{-8})  is consistent with the model predictions assuming a typical photo-desorption yield $Y_{\rm{H_2O}} = 1\times10^{-3}$ 
and a grain cross sectional area  $\sigma_{\rm{H}} = 2\times10^{-21}$\,\cmq. Indeed, the \citet{hollenbach09} model predicts that freeze-out and photo-desorption are the dominant processes to 
explain the water abundance starting at $A_{\rm V}\gtrsim 5$ for the physical conditions of the Mon~R2 envelope.

Summarizing, our observational results agree well with gas-phase chemical models only in the very first layers of the PDR, $A_{\rm V}\lesssim 1$, where high-temperature chemistry 
dominates the formation and photo-dissociation is the main destruction mechanism.  
The water abundance in the high-velocity gas is consistent with pure gas-phase chemistry, characteristic for PDRs. In the more quiescent gas, desorption from grains must play a role. This could be provided by photo-desorption according to the PDR model of \citet{hollenbach09}, by turbulence, or by grain destruction in shocks.
Deeper into the cloud, freeze-out into grains needs also to be taken into account for taking out \wat molecules from the gas phase.  
A complete treatment of the gas-grain chemistry is required to explain the observational results.

\subsection{Ammonia abundances}
Figure \ref{pdrmeudon} shows the predicted abundance for \amm as a function of the distance from the cloud center: qualitatively, ammonia is expected to be highly abundant only in the more  shielded part of the envelope.  This is consistent with our observational results, which show that emission of o-\amm is very extended, peaks outside the \HII region and  does not present high-velocity wings. This is also consistent with the results of the LVG modeling presented in Sect. \ref{sec_results}, which were obtained with  a beam filling factor of 1,  
densities between \ex{5}{4}\,\cmc and \ex{1}{5}\,\cmc and temperatures of $\sim 50$\,K. Lower densities would result in higher \amm abundances.
Comparing the abundances in the envelope obtained with the LVG modeling (few $10^{-9}$  for $n_{\rm{H_2}} = \ex{5}{4}$\,\cmc) 
 and with the Meudon PDR code (10$^{-8}$), it seems that gas-phase chemistry slightly over-predicts the observed values, although for less than an order of magnitude. 

\section{Summary and conclusions}
\label{sec_conclusions}

  In this work, we have presented spatially and spectrally resolved observations of [\CII], CO, o-\wat, p-\wat and \amm along a strip crossing the PDR that surrounds the UC\HII region MonR2. Using a local and  non-local radiative transfer model and assuming a spherical approximation for the whole cloud, we have fitted  the line profiles and intensities of the CO, \wat and \amm lines observed with \herschel.  
 
The o-H$_2^{18}$O line has a narrow profile that peaks at $\sim11$\,\kms, the rest velocity of the cloud.  The emission of this line is not very extended, and reaches it maximum toward the \HII region, suggesting that there is a significant fraction of the water emission arising from relatively quiescent gas associated to the innermost parts of the cloud. This is consistent with our modeling, which yielded an abundance of water of $\sim 1-3\times 10^{-8}$ relative to H$_2$ in a thick layer of 10 mag surrounding the PDR and in the envelope. 

The o-\wat abundance is higher, $\sim 10^{-7}$ relative to H$_2$, in red high-velocity gas. The origin of the red wings that appear in the \wat and in the high-J \dco lines 
is more doubtful. The red wings of all species toward the \HII region can be well reproduced by assuming an expanding PDR with relatively high expansion and turbulent 
velocities of 2\,\kms and 5\,\kms, respectively. Yet, the red wings could also have an outflow origin, particularly to the southwest half of the strip. 
In any case, the density of this gas is not very well constrained and therefore the precise values of the abundance 
need to be taken with caution, at an order-of-magnitude significance.

More work still remains to be done on the abundance of water in PDRs.  The comparison of the derived abundances with gas-phase chemical modeling shows that  gas-phase PDR chemistry can reproduce the observed abundances in the very first hot layer of the PDR up to an $A_{\rm V} \lesssim 1$. In this innermost layer, 
 high-temperature chemistry is the main source of production of gas-phase water  in the innermost layers of the PDR. However, the outflow can contribute to the \wat emission at these high velocities.		
In the more shielded layers of the molecular cloud, the observational abundances differ by several orders of magnitude from the predictions of the gas-phase chemical model, suggesting that the freeze-out and photo-desorption mechanisms are the dominant processes at these cloud depths. Similarly, gas-phase chemical models overpredict the abundance of ammonia in the envelope by about one order of magnitude, suggesting that freeze-out into the grain surface needs to be taken into account for this species.

\begin{acknowledgements}
 We acknowledge F. van der Tak for useful suggestions in improving this manuscript. We also acknowledge S. Trevi\~no-Morales and A. Sanchez-Monge for their help with complementary data.  We also thank the anonymous referee for his constructive report.\\
 This paper was partially supported by Spanish MICINN
under projects  and AYA2006-14876, AYA2009-07304 and within the program CONSOLIDER INGENIO 2010, under grant CSD2009-00038 Molecular Astrophysics: 
The Herschel and ALMA Era (ASTROMOL).\\
Part of this work was supported by the Deutsche
Forschungsgemeinschaft, project number Os 177/11.
HIFI has been designed and built by a consortium of institutes and university departments from across Europe, Canada and the United States under the leadership of SRON Netherlands Institute for Space Research, Groningen, The Netherlands and with major contributions from Germany, France and the US. Consortium members are: Canada: CSA, U.Waterloo; France: CESR, LAB, LERMA, IRAM; Germany: KOSMA, MPIfR, MPS; Ireland, NUI Maynooth; Italy: ASI, IFSI-INAF, Osservatorio Astrofisico di Arcetri-INAF; Netherlands: SRON, TUD; Poland: CAMK, CBK; Spain: Observatorio Astron—mico Nacional (IGN), Centro de Astrobiolog'a (CSIC-INTA). Sweden: Chalmers University of Technology - MC2, RSS \& GARD; Onsala Space Observatory; Swedish National Space Board, Stockholm University - Stockholm Observatory; Switzerland: ETH Zurich, FHNW; USA: Caltech, JPL, NHSC. \\

\end{acknowledgements}

\bibliographystyle{aa}
\bibliography{biblio}

\end{document}

%% file: authors.tex
\author{
P.~Pilleri\inst{1,2},
A.~Fuente\inst{2}, 
J.~Cernicharo\inst{1},
V.~Ossenkopf\inst{3,4},
O.~Bern\'e\inst{5,6},
M.~Gerin\inst{7},
J.~Pety\inst{8},
J.R.~Goicoechea\inst{1},
J.R.~Rizzo\inst{1},
J.~Montillaud\inst{9},
M.~Gonz\'alez-Garc\'{\i}a\inst{10},
C.~Joblin\inst{5,6},
J.~Le Bourlot\inst{11},
F.~Le Petit\inst{11}, 
C.~Kramer\inst{10}
}

 \institute{%
Centro de Astrobiolog\'{\i}a (INTA-CSIC),
             Ctra. M-108, km.~4, E-28850 Torrej\'on de Ardoz, Spain
\and
Observatorio Astron\'omico Nacional, Apdo. 112, E-28803 Alcal\'a de Henares, Spain			
\and
 I. Physikalisches Institut der Universit\"at 
 zu K\"oln, Z\"ulpicher Stra\ss{}e 77, 50937 K\"oln, Germany
\and
 SRON Netherlands Institute for Space Research, P.O. Box 800, 9700 AV 
 Groningen, Netherlands
 \and
Universit\'e de Toulouse; UPS-OMP; IRAP;  Toulouse, France
 \and
CNRS; IRAP; 9 Av. colonel Roche, BP 44346, F-31028 Toulouse cedex 4, France 
\and
LERMA, Observatoire de Paris, 61 Av. de l'Observatoire, 75014 Paris, France 
\and
Institut de Radioastronomie Millim\'etrique, 300 Rue de la Piscine, 38406 Saint Martin d'H\'eres, France
 \and
Department of Physics, P.O.Box 64, FI-00014, University of Helsinki, Finland
\and
 Instituto de Radio Astronom\'ia Milim\'etrica (IRAM), Avenida Divina Pastora 7, Local 20, 18012 Granada, Spain
\and
Observatoire de Paris, LUTH and Universit\'e Denis Diderot, Place J.
Janssen, 92190 Meudon, France
}

\offprints{P. Pilleri, \email{p.pilleri@oan.es}}